\documentclass[12pt,letterpaper]{article}
\usepackage[top=0.85in,left=0.85in,footskip=0.75in,marginparwidth=1in]{geometry}
\usepackage{amsmath}
\usepackage{upgreek}
\usepackage{xcolor}
\usepackage{graphicx}
\usepackage{upgreek}

\usepackage[utf8]{inputenc}

\usepackage{cite}

\usepackage{nameref,hyperref}

\usepackage{microtype}
\DisableLigatures[f]{encoding = *, family = * }

\raggedright
\usepackage{changepage}

\usepackage[aboveskip=1pt,labelfont=bf,labelsep=period,singlelinecheck=off]{caption}

\makeatletter
\renewcommand{\@biblabel}[1]{\quad#1.}
\makeatother

\usepackage{lastpage,fancyhdr,graphicx}
\usepackage{epstopdf}
\fancyheadoffset[L]{2.25in}
\fancyfootoffset[L]{2.25in}

\usepackage{color}

\definecolor{Gray}{gray}{.25}

\usepackage{graphicx}

\usepackage{sidecap}

\usepackage{wrapfig,cite}
\usepackage[pscoord]{eso-pic}
\usepackage[fulladjust]{marginnote}
\reversemarginpar

\usepackage{ragged2e}
\justifying\let\raggedright\justifying

\begin{document}

	\vspace*{0.35in}
	
	\begin{flushleft}
		{\Large
			\textbf\newline{Realization of broadband truly rainbow trapping in gradient-index heterostructures}
		}
		\newline
		
		Jie Xu,\textsuperscript{1,2} Sanshui Xiao,\textsuperscript{3} Panpan He,\textsuperscript{4} Yazhou Wang,\textsuperscript{3} Yun Shen,\textsuperscript{5} Lujun Hong,\textsuperscript{5,7} Yamei Luo,\textsuperscript{1,2,6} and Bing He,\textsuperscript{1,2,8}
		
		\bigskip
		\bf{1} School of Medical Information and Engineering, Southwest Medical University, Luzhou 646000, China\\
		\bf{2} Medicine \& Engineering \& Information Fusion and Transformation Key Laboratory of Luzhou City, Luzhou 646000, China\\
		\bf{3} DTU Fotonik, Department of Photonics Engineering, Technical University of Denmark, DK-2800 Kgs. Lyngby, Denmark\\
		\bf{4} Department of Electronic Engineering, Luzhou Vocational and Technical College, Luzhou 646000, China\\
		\bf{5} Institute of Space Science and Technology, Nanchang University, Nanchang 330031, China\\
		\bf{6} luoluoeryan@126.com\\
		\bf{7} ljhong@ncu.edu.cn\\
		\bf{8} hebing\_wu@163.com\\


	\end{flushleft}
	
	\section*{Abstract}
	Unidirectionally propagating waves (UPW) such as topologically protected edge modes and surface magnetoplasmons (SMPs) has been a research hotspot in the last decades. In the study of UPW, metals are usually treated as perfect electric conductors (PECs) which, in general, are the boundary conditions. However, it was reported that the transverse resonance condition induced by the PEC wall(s) may significantly narrow up the complete one-way propagation (COWP) band. In this paper, we propose two ways to achieve ultra-broadband one-way waveguide in terahertz regime. The first way is utilizing the epsilon negative (ENG) metamaterial (MM) and the other one is replacing the PEC boundary with perfect magnetic conductor (PMC) boundary. In both conditions, the total bandwidth of the COWP bands can be efficiently broadened by more than three times. Moreover, based on the ultra-broadband one-way configurations, gradient-index metamaterial-based one-way waveguides are proposed to achieve broadband truly rainbow trapping (TRT). By utilizing the finite element method, the realization of the broadband TRT without backward reflection is verified in gradient-index structures. Besides, giant electric field enhancement is observed in a PMC-based one-way structure with an ultra-subwavelength ($\approx 10^{-4} \lambda_0$, $\lambda_0$ is the wavelength in vaccum) terminal, and the amplitude of the electric field is enormously enhanced by five orders of magnitude. Our findings are beneficial for researches on broadband terahertz communication, energy harvesting and strong-field devices.
	
	
	\section{Introduction}
	
	One-way or unidirectional electromagnetic (EM) modes attract more and more attentions in the past two decades\cite{Takeda:Co,Yu:On,Ao:On,Skirlo:Mu,Yang:Vi} for its unique unidirectional propagation property which has extensive applications in optical communication. Similar to chiral edge states in quantum Hall effect, the one-way EM waves are immune to the backscattering, which has been observed in experiments at microwave frequency\cite{Prang:Th,Wang:Ob}. Engineering the band diagram of photonic crystals (PhCs) by introducing disorders is an efficient way to build one-way waveguide\cite{Haldane:Po,Wang:Re,Yu:On}, and the cells of PhCs always consist of magneto-optical (MO) materials under a static magnetic field which is exploited to break the time-reversal symmetry of the system. Recently, numerous applications such as optical splitter\cite{Zhang:Co,Hong:Hi,He:Tu}, subwavelength focusing\cite{Shen:Tr,Xu:Ul,Shen:Co} and optical switch\cite{Hu:Br} were proposed based on the one-way waveguides. More recently, Tsakmakidis' group pointed out that the time-bandwidth (TB) limit which are believed to be a fundamental limit in engineering and physics, can be broken in a carefully designed one-way waveguide. In their work, Tsakmakidis et al. reported that the one-way surface magnetoplasmons (SMPs) propagating on the interface of silicon and InSb can be trapped and be localized in a subwavelength zero-dimensional cavity, and the TB limit was broken in such structure because of the nonreciprocity of such system\cite{Tsakmakidis:Br}.
	
	According to the special theory of relativity, the speed of light must be a constant which means the light cannot be accelerated. On the contrary, slowing light or EM waves is much more easier. In the past decades, slowed or even trapped light was found in ultracold atomic gas\cite{Hau:Li}, PhCs\cite{Baba:Sl,Schulz:Di,Gersen:Re,Yoshimi:Sl}, grating structures\cite{Gan:Ul,Gan:Ex}, metamaterials (MMs)\cite{Tsakmakidis:Tr,Zhang:Ab,Gao:Du} and MO material heterostructures\cite{Xu:Br,Yang:Ex}. Among the slow-light structures, only a few of them can trap EM waves with different frequencies at different locations, namely rainbow trapping. On the other hand, due to the coupling effect and reflection, most of the rainbow trapping systems cannot truly capture the EM waves\cite{Liu:Tr,He:Re}. The key point of truly rainbow trapping (TRT) is blocking the coupling between forward and backward propagating waves. Lately, we proposed a metal-semiconductor-semiconductor-metal structure, and by judiciously designing the thickness of the semiconductor layers, we achieved TRT in terahertz regime\cite{Xu:Sl}. 
	
	Nowadays, the study of the EM MMs with anomalous characteristics, without doubt, is one of the most interesting research directions. Novel MMs such as optical negative-index MMs\cite{Xiao:Lo,Shalaev:Op,Valentine:Th}, hyperbolic MMs\cite{Poddubny:Hy,Hu:Ra,Guo:Th}, artificial magnetic conductor (AMC) MMs\cite{Monti:Ac,Erentok:Ch} and epsilon-negative (ENG) MMs\cite{Moniruzzaman:Cr,Dawar:Ba,Kumari:An} have been reported. Moreover, epsilon-near-zero (ENZ) MMs attract many attentions for its potential uses in studies of large nonlinearity\cite{Alam:La}, supercoupling\cite{Edwards:Ex}, lens\cite{Torres:Te} and two dimensional materials\cite{Biswas:Tu}. Note that the ENZ MMs have been achieved using techniques such as effective medium theory\cite{Jason:Ex}, semiconductor doping\cite{Alam:La} and applying a static electric field\cite{Feigenbaum:Un}. However, the study of stable ENG or ENZ MMs is still a research gap and in our opinion, reconfiguring the EM properties of MMs by cladding liquid crystal\cite{Werner:Li} may be a possible way to achieve such stable ENG or ENZ MMs. In this paper, we consider the stable ENG and ENZ MMs in the study of ultra-broadband one-way waveguides and TRT. By utilizing the ENG/ENZ MMs and PMC wall(s), the complete one-way (unidirectional) propagation (COWP) band of the one-way configurations are significantly broadened. Moreover, different with our previous work\cite{Xu:Sl} where the TRT was achieved in tapered waveguides for a relative narrow band, we propose novel configurations consisting of gradient-index mediums in this work to achieve broadband TRT. Besides, we investigate the uses of perfect magnetic conductor (PMC) boundary in the one-way waveguides and interestingly, giant electric field enhancement is observed in a PMC-based one-way waveguide which has an ultra-subwavelength terminal. All the theoretical analysis are verified by simulations utilizing finite element method.
	
	\section{Ultra-broadband one-way waveguide and broadband TRT}
	\begin{figure}[ht]
		\centering\includegraphics[width=3.5 in]{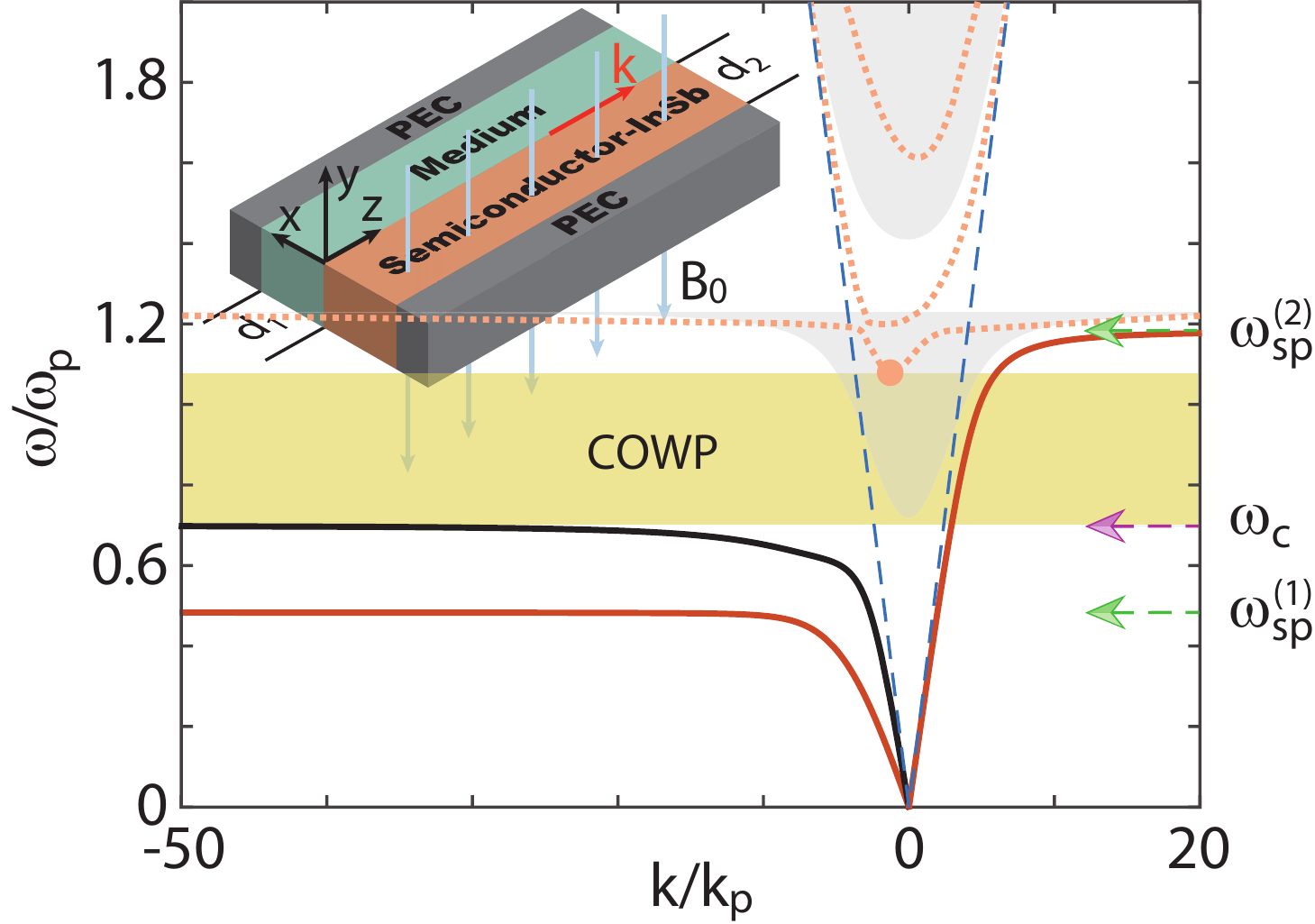}
		\caption{ The dispersion diagram of the PEC-Medium-Semiconductor-PEC (EMSE) model as $\omega_\mathrm{c}=0.7\omega_\mathrm{p}$, $d_1=d2=0.05\lambda_\mathrm{p}$ and $\varepsilon_\mathrm{m}=11.68$. The red and black curves represent SMPs and SMs, respectively. The dotted lines are the low orders of normal modes in the medium (silicon, Si) and InSb layers, while the gray shaded zones and the blue dashed lines represent the bulk zones in semi-infinite InSb and the light lines of Si. The inset shows the schematic of the EMSE structure.}\label{Fig1}
	\end{figure}
	
	We first investigate the dispersion characteristics in a typical one-way terahertz structure and, as shown the inset of Fig. 1, the physical model consists of two layers of metal which can be treated as perfect electric conductor (PEC) in terahertz regime, and one layer of medium and one layer of semiconductor. We note here that, in this paper, the semiconductor is assumed to be N-type InSb. An external magnetic field ($B_0$) is applied on the InSb layer to break the time-reversal symmetry in the system. In our previous works, we have proved that the loss effect has almost no impact on the one-way property in such MO system\cite{Shen:On,Xu:Sl}. Thus, we study the dispersion relation of SMPs in lossless condition in this work. The permittivity of InSb, as reported in many works\cite{Xu:Br,Shen:On,Tsakmakidis:Br}, can be written as follow
	\begin{equation}
		\stackrel{\leftrightarrow}{\varepsilon}=\left[\begin{array}{ccc}
			\varepsilon_{1} & 0 & i \varepsilon_{2} \\
			0 & \varepsilon_{3} & 0 \\
			-i \varepsilon_{2} & 0 & \varepsilon_{1}
		\end{array}\right]
	\end{equation}
	with
	\begin{equation}
		\varepsilon_{1}=\varepsilon_{\infty}\left(1-\frac{\omega_{\mathrm{p}}^{2}}{\omega^{2}-\omega_{\mathrm{c}}^{2}}\right), \varepsilon_{2}=\varepsilon_{\infty} \frac{\omega_{\mathrm{c}} \omega_{\mathrm{p}}^{2}}{\omega\left(\omega^{2}-\omega_{\mathrm{c}}^{2}\right)}, \varepsilon_{3}=\varepsilon_{\infty}\left(1-\frac{\omega_{\mathrm{p}}^{2}}{\omega^{2}}\right),
	\end{equation}
	where $\varepsilon_\infty$, $\omega$, $\omega_\mathrm{p}$, $\omega_\mathrm{c}=eB_0/{m^*}$ (e and $m^*$ represent the charge and the effective mass of an electron) are the high-frequency permittivity, the angular frequency, the plasma frequency and the electron cyclotron frequency, respectively. In this PEC-medium-semiconductor-PEC (EMSE) structure, the SMPs are sustained by the medium-InSb interface and due to the transverse resonance condition induced by the PEC walls, the dispersion relation of the SMPs should be distinctly changed compared to the one in the medium-semiconductor (MS) waveguide. The dispersion equation of the surface EM modes can be easily obtained by combining the Maxwell's equations and the boundary conditions, and according to our calculation, it has the following form\cite{Xu:Br}
	\begin{equation}
		\left(k^{2}-\varepsilon_{1} k_{0}^{2}\right) \tanh \left(\alpha d_{2}\right)+\frac{\varepsilon_{1}}{\varepsilon_\mathrm{m}} \alpha_\mathrm{m}\left[\alpha-\frac{\varepsilon_{2}}{\varepsilon_{1}} k \tanh \left(\alpha d_{2}\right)\right] \tanh \left(\alpha_\mathrm{m} d_{1}\right)=0
	\end{equation}
	where $\alpha=\sqrt{k^2-\varepsilon_v k_0^2}$ ($\varepsilon_v=\varepsilon_1-\varepsilon_2^2/\varepsilon_1$ is the Voigt permittivity), $\alpha_\mathrm{m}=\sqrt{k^2-\varepsilon_\mathrm{m} k_0^2}$ ($\varepsilon_\mathrm{m}$ is the relative permittivity of medium) represent the transverse attenuation coefficients in the InSb layer and in the medium layer, respectively. $d_1$ and $d_2$ are the thicknesses of the medium and the InSb layers. Since we already have the dispersion equation of the SMPs, we can easily obtain the values of the asymptotic frequencies (AFs), one of the most important characteristics in the study of (nonreciporcal) unidirectional waveguides. Interestingly, when $k\rightarrow \pm \infty$, three asymptotic frequencies (AFs) are found in Eq. (3) and they are written as below
	\begin{equation}
		\omega_\mathrm{sp}^{(1)}=\frac{1}{2}\left(\sqrt{\omega_\mathrm{c}^{2}+4 \frac{\varepsilon_{\infty}}{\varepsilon_{\infty}+\varepsilon_\mathrm{m}} \omega_\mathrm{p}^{2}}-\omega_\mathrm{c}\right), \omega_\mathrm{sp}^{(2)}=\frac{1}{2}\left(\sqrt{\omega_\mathrm{c}^{2}+4 \frac{\varepsilon_{\infty}}{\varepsilon_{\infty}+\varepsilon_\mathrm{m}} \omega_\mathrm{p}^{2}}+\omega_\mathrm{c}\right), 
		\omega_\mathrm{sp}^{(3)}=\omega_\mathrm{c}
	\end{equation}
	To clearly show the special three AFs property, we plot the dispersion diagram of the EM modes in the EMSE waveguide in Fig. 1 for $\omega_\mathrm{c}=0.7\omega_\mathrm{p}$, $d_1=d_2=0.05\lambda_\mathrm{p}$ ($\lambda_\mathrm{p}=2\pi c/\omega_\mathrm{p}$) and $\varepsilon_\mathrm{m}=11.68$ (i.e. silicon). We note here that we have investigated the EMSE structure in previous work, however, in this work, we will further explore the capability of the EMSE in building ultra-broadband one-way structures by utilizing MMs such as ENG MMs. In Fig. 1, the red solid lines, the black line and the dotted lines indicate the dispersion curves of SMPs and SMs\cite{Xu:Sl} sustained by the semiconductor-metal interface and lowest-order normal modes in the light cone or bulk zones (the gray shaded regions), respectively. Note that the dispersion relation of the SMs and the normal modes can be easily derived from the Eq. (3) by changing the real and imaginary properties of $\alpha$ and $\alpha_\mathrm{m}$. Moreover, three horizontal arrows in Fig. 1 represent the values of three AFs. The yellow shaded region is the complete one-way propagation (COWP) band which is confined by $\omega_\mathrm{c}$ ($=0.7\omega_\mathrm{p}$) and the cut-off frequency ($\omega_\mathrm{cf} \approx 1.079\omega_\mathrm{p}$, marked by the orange point) of the lowest-order normal modes. It is clear that in the $\omega<\omega_\mathrm{c}$ region, the EM modes can propagate in both forward and backward directions since there are SMs with $v_\mathrm{g}<0$ and SMPs with $v_\mathrm{g}>0$. As a consequence, the bandwidth ($\Delta \omega \approx 0.379\omega_\mathrm{p}$) of the COWP band is relatively small in this case. A possible way to widen the COWP band in such structure is to reduce or even eliminate the influences of SMPs in the $\omega<\omega_\mathrm{c}$ region. In what follows, we will show that the ENG MMs could be used to achieve ultra-broadband one-way waveguide. 
	\begin{figure}[pt]
		\centering\includegraphics[width=4.5 in]{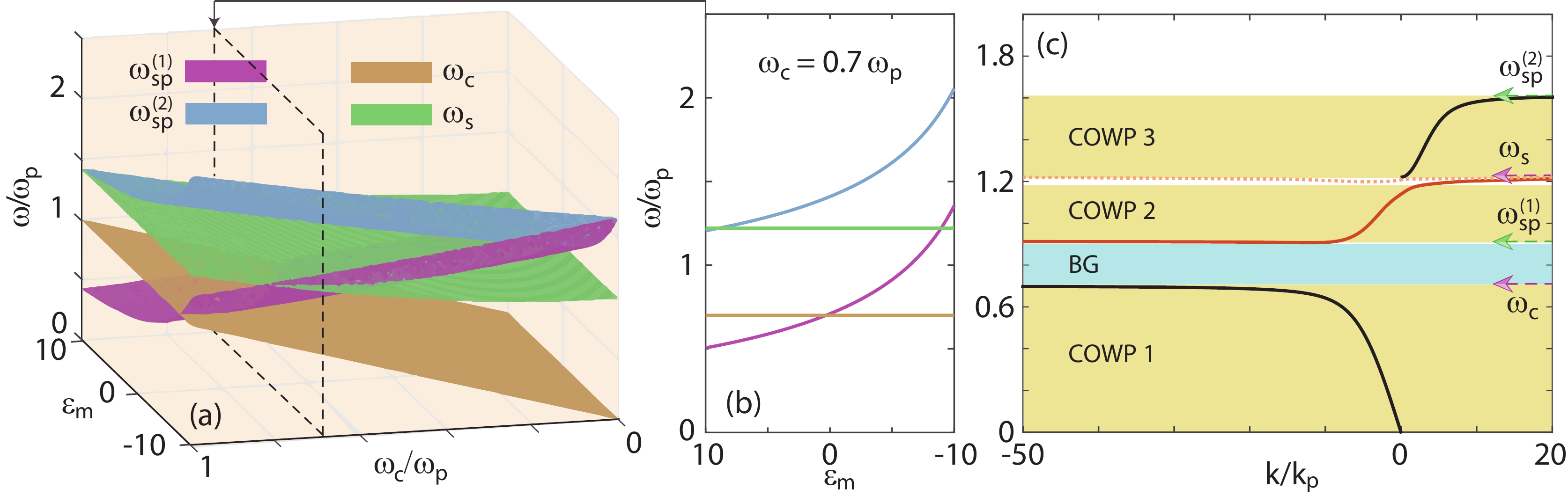}
		\caption{ $\omega_\mathrm{sp}^{(1)}$, $\omega_\mathrm{sp}^{(2)}$, $\omega_\mathrm{c}$ and $\omega_\mathrm{s}$ as a function of $\varepsilon_\mathrm{m}$ as (a) $0<\omega_\mathrm{c}<1$ and (b) $\omega_\mathrm{c}=0.7\omega_\mathrm{p}$. (c) The dispersion diagram of surface EM modes in Ultra-broadband one-way waveguide with $\varepsilon_\mathrm{m}=-5$. The other parameters are the same as in Fig. 1.}\label{Fig2}
	\end{figure}
	
	As shown in Fig. 2(a), the values of $\omega_\mathrm{sp}^{(1)}$, $\omega_\mathrm{sp}^{(2)}$, $\omega_\mathrm{c}$ and $\omega_\mathrm{s}$ ($\omega_\mathrm{s}=\sqrt{\omega_\mathrm{c}^2+\omega_\mathrm{p}^2}$) were plotted as a function of $\varepsilon_\mathrm{m}$ and $\omega_\mathrm{c}$. As a result, $\omega_\mathrm{c}$ (the brown surface) is always smaller than $\omega_\mathrm{s}$ (the green surface) while $\omega_\mathrm{sp}^{(1)}$ (the purple surface) is invariably smaller than $\omega_\mathrm{sp}^{(2)}$ (the blue surface). More interestingly, as shown in Fig. 2(b), $\omega_\mathrm{sp}^{(1)}>\omega_\mathrm{c}$ holds for $\varepsilon_\mathrm{m}<0$ as $\omega_\mathrm{c}=0.7$, implying that the dispersion curves of the SMPs may rise up in the case of $\varepsilon_\mathrm{m}<0$. Figure 2(c) illustrates the dispersion diagram of the EM modes in the EMSE model with $\varepsilon_\mathrm{m}=-5$. Compared to Fig. 1, Fig. 2(c) demonstrates four AFs, two of which are $\omega_\mathrm{sp}^{(1)}$ and $\omega_\mathrm{sp}^{(2)}$ and they are significantly increased compared to the ones in Fig. 1. More importantly, the SMPs' dispersion curves rose up into the $\omega>\omega_\mathrm{c}$ band, leading to three COWP bands (the yellow shaded regions) and one band gap (the cyan region). The total bandwidth $\Delta \omega$ ($\approx 1.396\omega_\mathrm{p}$) of the COWP regions in this case is about 3.7 times of the one in Fig. 1. In addition, the completely disappeared SMPs in the $0<\omega<\omega_\mathrm{c}$ band implies that the ENG MMs-based MO heterostructures have potential to excite pure one-way propagating SMs, which should have further uses in exploring on-chip nonreciprocal plasmonics\cite{Liang:Tu}.
	
	\begin{figure}[pt]
		\centering\includegraphics[width=4.5 in]{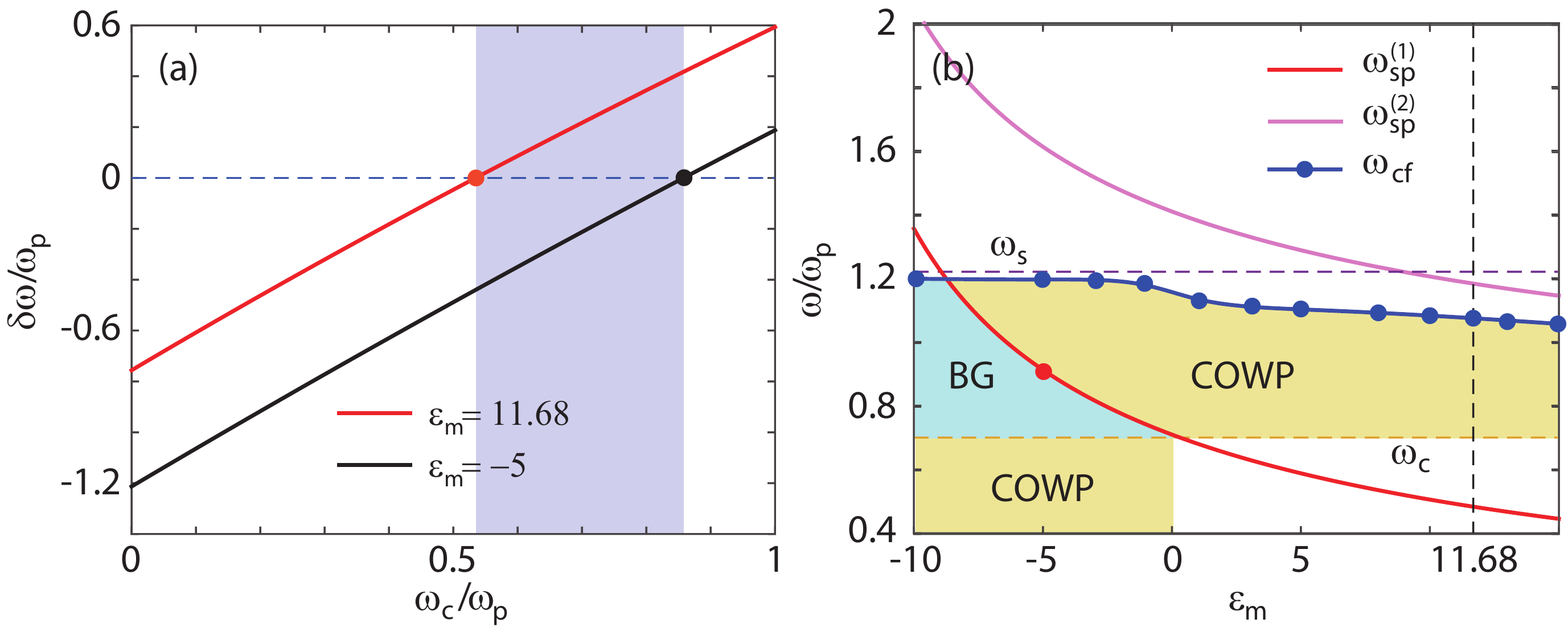}
		\caption{ (a) The relation between $\delta \omega =\omega_\mathrm{c}-\omega_\mathrm{sp}^{(1)}$ and $\omega_\mathrm{c}$ in Si- and ENG-based EMSE waveguides for $0<\omega_\mathrm{c}<\omega_\mathrm{p}$. (b) Three asymptotic frequencies, i.e. $\omega_\mathrm{sp}^{(1)}$, $\omega_\mathrm{sp}^{(2)}$ and $\omega_\mathrm{c}$, and the cut-off frequencies $\omega_{cf}$ as a function of $\varepsilon_\mathrm{m}$.}\label{Fig3}
	\end{figure}	
	The emerged band gap (BG) in Fig. 2(c) is limited by $\omega_\mathrm{c}$ and the cut-off frequency ($\approx \omega_\mathrm{sp}^{(1)}$) of the dispersion branch of the SMPs (the red line), and excitingly, the BG falls in the COWP band in Fig. 1. On the other hand, it is clear that $\omega_\mathrm{sp}^{(1)}$ gradually increases as $\varepsilon_\mathrm{m}$ decreases whereas $\omega_\mathrm{c}$ is a constant. Therefore, the BG should be widen when decreasing $\varepsilon_\mathrm{m}$. Accordingly, we believe that the truly rainbow trapping (TRT) can be achieved in novel heterostructures where the permittivity of the medium linearly decreases along the propagation direction ($+z$). To verify our conjecture, in Fig. 3(a), we first plot the values of $\delta \omega = \omega_\mathrm{c}-\omega_\mathrm{sp}^{(1)}$ as a function of $\omega_\mathrm{c}$ for $\varepsilon_\mathrm{m}=11.68$ and $\varepsilon_\mathrm{m}=-5$. The shaded area show the region in which $\delta \omega>0$ for $\varepsilon_\mathrm{m}=11.68$ and $\delta \omega<0$ for $\varepsilon_\mathrm{m}=-5$. To achieve the TRT in the one-way waveguides consisting of medium with $-5\leq\varepsilon_\mathrm{m}\leq11.68$, the external magnetic field should have a appropriate value and $\omega_\mathrm{c}$ should fall in the shaded region, i.e. $0.5347<\omega_\mathrm{c}<0.8578\omega_\mathrm{p}$. As an example, we set $\omega_\mathrm{c}=0.7\omega_\mathrm{p}$ and the broadband TRT theory are illustrated in Figure 3(b), in which $\omega_\mathrm{sp}^{(1)}$ (the red line), $\omega_\mathrm{sp}^{(2)}$ (the pink line), $\omega_\mathrm{cf}$ (the blue points) are plotted as a function of $\varepsilon_\mathrm{m}$. The yellow regions and the cyan region represent the COWP bands and the BG, respectively. It is clear that, in Fig. 3(b), there is usually only one COWP band for $\varepsilon_\mathrm{m}>0$, and there are one or two (depending on $\varepsilon_\mathrm{m}$) COWP bands and one BG for $\varepsilon_\mathrm{m}<0$. Note that with the decreases of $\varepsilon_\mathrm{m}$, part of the COWP band gradually changes to a BG. The red point represents $\omega \approx 0.9126\omega_\mathrm{p}$ which is the upper limit of the BG in the case of $\varepsilon_\mathrm{m}=-5$. Meanwhile, the BG ($0.7\omega_\mathrm{p} \leq \omega \leq 0.9126 \omega_\mathrm{p}$) falls in the COWP band in the case of $\varepsilon_\mathrm{m}=11.68$ (the black dashed line). Therefore, one can easily conclude that the EM wave with working frequency falls in $(0.7\omega_\mathrm{p},0.9126\omega_\mathrm{p})$ band can unidirectionally propagating at the position of $\varepsilon_\mathrm{m}=11.68$ and be forbidden when $\varepsilon_\mathrm{m}\leq-5$.
	
	\begin{figure}[ht]
		\centering\includegraphics[width=4 in]{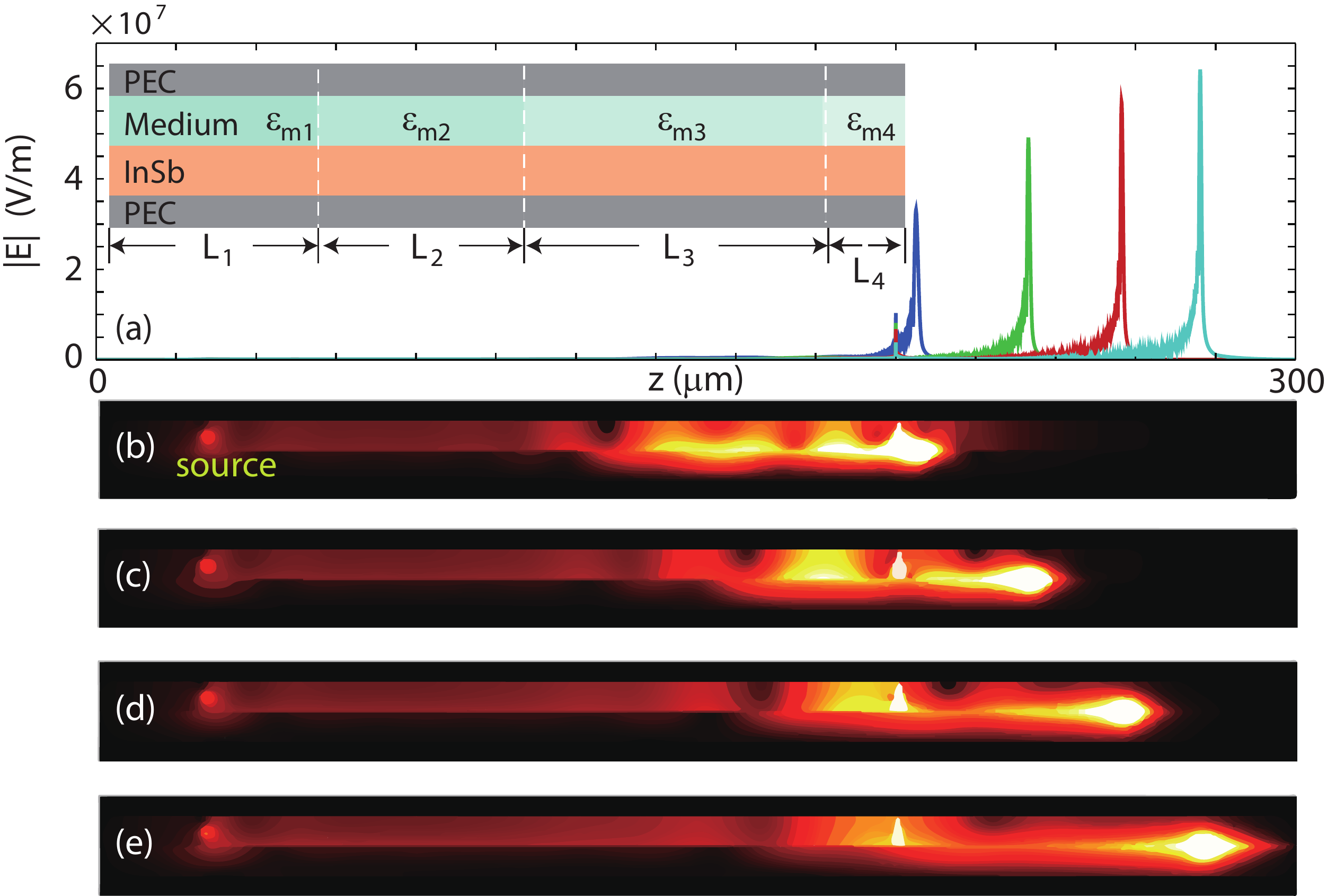}
		\caption{ (a) The electric field amplitude along the medium-InSb interface in the FEM simulations. The inset illustrates the designed rainbow-trapping EMSE configuration. (b)-(e) The simulated electric field distributions for $f=0.72f_\mathrm{p}$, $f=0.78f_\mathrm{p}$, $f=0.84f_\mathrm{p}$ and $f=0.9f_\mathrm{p}$. The electron scattering frequency are considered to be $\nu = 0.001\omega_\mathrm{p}$ in the FEM simulations.}\label{Fig4}
	\end{figure}
	
	Based on the results demonstrated in Fig. 3, we designed a novel EMSE waveguide (see the inset of Fig. 4(a)) which has four different EMSE parts, and the lengths of the four parts are respectively $L_1=80\upmu$m, $L_2=40\upmu$m, $L_3=160\upmu$m and $L_4=20\upmu$m. Note that, in the simulations performed in this paper, the loss tangent of the ENG MMs was assumed to be 0.001 and the loss effects of InSb layer were considered by introducing the scattering frequency $\nu=0.001\omega_\mathrm{p}$, and the other parameters of InSb are $\varepsilon_\infty=15.6$ and $\omega_\mathrm{p}=4\pi\times10^{12}$ rad/s\cite{Isaac:De}. Different with regular one-way waveguides, we assume the permittivities of the four parts of the EMSE struture are $\varepsilon_\mathrm{m1}=11.68$, $\varepsilon_\mathrm{m2}=11.68-6.68(z-L_1)/L_2$, $\varepsilon_\mathrm{m3}=5-10(z-(L_1+L_2))/L3$ and $\varepsilon_\mathrm{m4}=-5$, respectively. It is worth noting that the gradient-index medium has been proposed in configurations such as sonic crystal\cite{Climente:So} and photonic crystal\cite{Centeno:Mi}. The thicknesses of InSb and mediums are the same as in Fig. 3(b). Based on the above analysis, we know that $\omega_\mathrm{sp}^{(1)}\approx0.9126\omega_\mathrm{p}$ for $\varepsilon_\mathrm{m}=-5$, and, furthermore, the EM waves with $0.7\omega_\mathrm{p}<\omega<0.9126\omega_\mathrm{p}$ should be trapped in our proposed waveguide shown in the inset of Fig. 4(a). Fig. 4(b)-4(f) shows the electric field distributions in the simulations using finite element method (FEM) and the working frequencies are respectively $f=0.72f_\mathrm{p}$ ($f_\mathrm{p}=2$ $\mathrm{THz}$), $f=0.78f_\mathrm{p}$, $f=0.84f_\mathrm{p}$ and $f=0.9f_\mathrm{p}$. As a result, the EM waves with different frequencies are truly trapped in different place in the waveguide, which means the TRT are achieved in such structure. Here we emphasis that only the trapped rainbow without back reflection can be called truly trapped rainbow\cite{Xu:Sl,Liu:Tr}. Moreover, the EM waves in four simulations present the property of one-way propagation which is the most difference between regular rainbow trapping and one-way SMPs-based TRT. Fig. 4(a) shows the amplitudes of the electric field on the medium-InSb interface and the phenomenons of rainbow trapping and electric field enhancement were observed at the same time. Similarly, according to Fig. 3(b), it is capable of achieving the broadband TRT with the bandwidth $\Delta \omega_\mathrm{TRT} \approx 0.4\omega_\mathrm{p}$ in such gradient-index heterostructure. We emphasize that the above $\Delta \omega_\mathrm{TRT}$ is more than three times wider than the one reported in our previous work\cite{Xu:Sl}. Besides the ENG MMs, in the next subsection, we will show that the PMC wall(s) can be exploited to achieve the broadband TRT as well.

	\section{PMC-based one-way waveguides}
	\begin{figure}[ht]
		\centering\includegraphics[width=4.5 in]{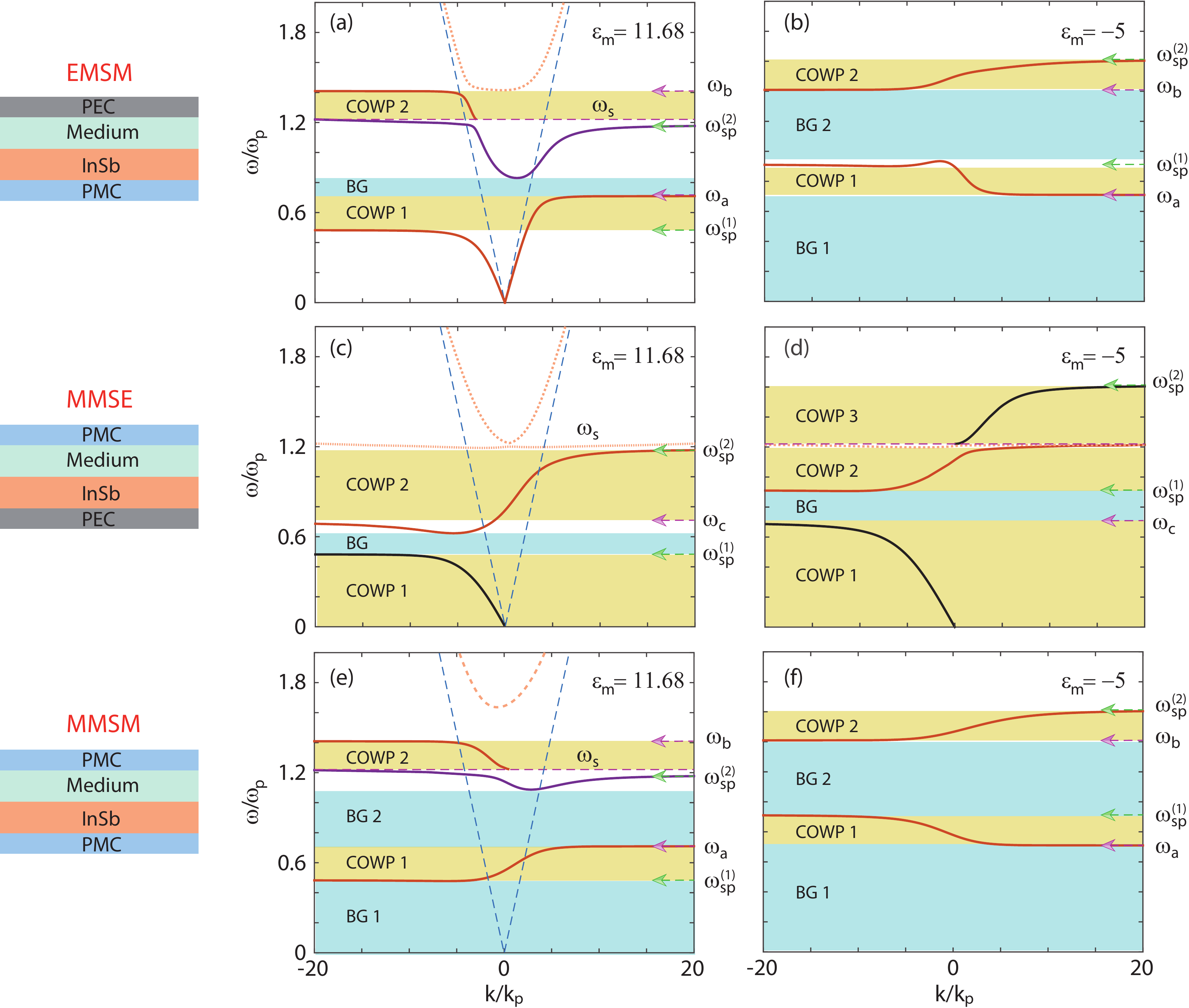}
		\caption{ The dispersion diagram of (a,b) the EMSM, (c,d) the MMSE and (e,f) the MMSM configurations. The mediums are considered being (a,c,e) silicon and (b,d,f) ENG MMs with $\varepsilon_\mathrm{m}=-5$.}\label{Fig5}
	\end{figure}
	\begin{figure}[ht]
		\centering\includegraphics[width=4.5 in]{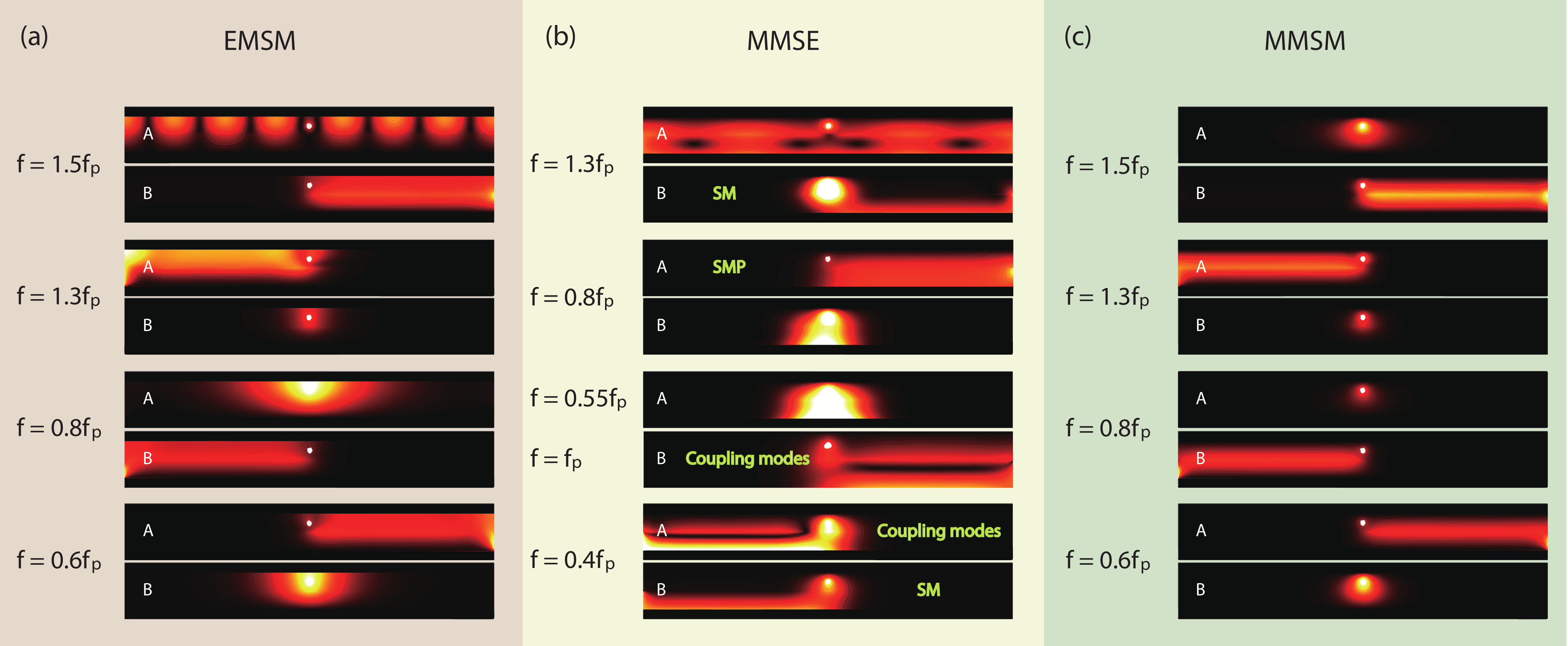}
		\caption{ The magnetic field distributions of the FEM simulations of (a) the EMSM, (b) the MMSE and (c) the MMSM structures. The other parameters are the same as in Fig. 5.}\label{Fig6}
	\end{figure}
	\begin{figure}[ht]
		\centering\includegraphics[width=4 in]{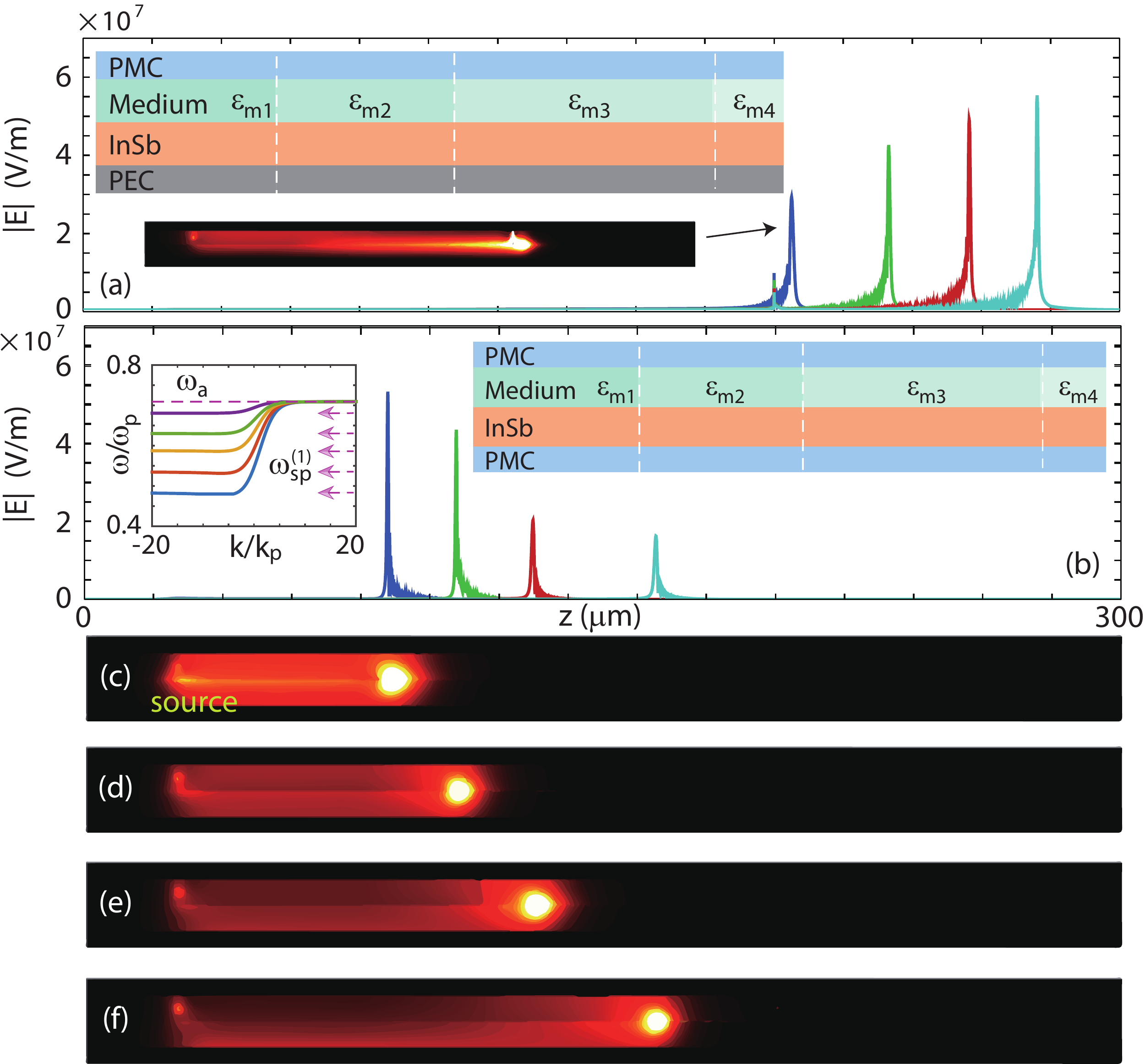}
		\caption{ Similar TRT as in Fig. 4 based on (a) the MMSE and (b) the MMSM configurations. The lower inset of (a) shows one of FEM simulation results of the trapped rainbow as $f=0.72f_\mathrm{p}$. The left inset of (b) represents the dispersion curves of the MMSM structure as $\varepsilon_\mathrm{m}=11.68$ (blue line), $\varepsilon_\mathrm{m}=8$ (red line), $\varepsilon_\mathrm{m}=5$ (yellow line), $\varepsilon_\mathrm{m}=3$ (green line) and $\varepsilon_\mathrm{m}=1$ (purple line). The schematics of the MMSE and MMSM structures shown in (a) and (b) are almost the same with the one in Fig. 4(a) except for one or two layers PEC walls are replaced by PMC walls. (c)-(f) The electric field distributions in TRT based on the MMSM waveguide. Four working frequencies are $0.5f_\mathrm{p}$, $0.55f_\mathrm{p}$, $0.6f_\mathrm{p}$ and $0.65f_\mathrm{p}$, respectively.}\label{Fig7}
	\end{figure}
	Metal layers are usually treated as PEC walls in terahertz regime and based on our previous works, we found that the existence of the PEC walls may destroy the property of one-way propagation in ultra-subwavelength waveguides for the transverse resonance condition is introduced in such structures\cite{Xu:Br,Xu:Sl}. Here, we emphasize that the metal layers (PEC walls) can be replaced by PMC walls to break the limitation in the study of one-way EM modes in ultra-subwavelength MO heterostructures. As shown the left diagrams of Fig. 5, there are three ways using PMC wall(s), i.e. replacing the lower PEC wall, replacing the upper PEC wall and replacing all the PEC walls with PMC wall(s) in the EMSE configuration. Three new structures are respectively the PEC-medium-semiconductor-PMC (EMSM) structure, the PMC-medium-semiconductor-PEC (MMSE) structure and the PMC-medium-semiconductor-PMC (MMSM) structure. Similarly, the dispersion equations of the surface modes in these waveguides can be written as follow
	\begin{equation}
		\frac{\varepsilon_{2}}{\varepsilon_{1}} k + \frac{\alpha}{\tanh (\alpha d_2)}+
		\frac{\varepsilon_v}{\varepsilon_\mathrm{m}}\alpha \tanh \left(\alpha_\mathrm{m} d_{1}\right)=0, \qquad (\mathrm{EMSM})
	\end{equation}
	\begin{equation}
		\left(k^{2}-\varepsilon_{1} k_{0}^{2}\right) \tanh \left(\alpha d_{2}\right)+\frac{\varepsilon_{1}}{\varepsilon_\mathrm{m}} \alpha_\mathrm{m}\left[\alpha-\frac{\varepsilon_{2}}{\varepsilon_{1}} k \tanh \left(\alpha d_{2}\right)\right] \frac{1}{\tanh \left(\alpha_\mathrm{m} d_{1}\right)}=0, \qquad (\mathrm{MMSE})
	\end{equation}
	\begin{equation}
		\frac{\varepsilon_{2}}{\varepsilon_{1}} k + \frac{\alpha}{\tanh (\alpha d_2)}+ \frac{\varepsilon_v}{\varepsilon_\mathrm{m}}\frac{\alpha}{\tanh \left(\alpha_\mathrm{m} d_{1}\right)}=0. \qquad (\mathrm{MMSM})
	\end{equation}
	We further plot the diagrams of the dispersion curves of the EM waves in these structures in Fig. 5 as $\omega_\mathrm{c}=0.7\omega_\mathrm{p}$ and $d_1=d_2=0.05\lambda_\mathrm{p}$ ($\lambda_\mathrm{p}=150$ $\upmu$m). Si and the ENG MM with $\varepsilon_\mathrm{m}=-5$ are considered as two kinds of mediums in the EMSM waveguide (see Figs. 5(a,b)), in the MMSE waveguide (see Figs. 5(c,d)) and in the MMSM waveguide (see Figs. 5(e,f)). From Fig. 5, one can see that, for $\varepsilon_\mathrm{m}=11.68$, the numbers ($N_\mathrm{si}$) of the COWP band ($N_\mathrm{cowp}$) and the BG ($N_{bg}$) in the EMSM, MMSE and MMSM structures are $N_\mathrm{si}=(N_\mathrm{cowp},N_\mathrm{bg})=(2,1)$, $N_\mathrm{si}=(2,1)$ and $N_\mathrm{si}=(2,2)$, respectively. On the other hand, the corresponding numbers ($N_\mathrm{eng}$) of the COWP band and the BG in the cases of $\varepsilon_\mathrm{m}=-5$ are $N_\mathrm{eng}=(2,2)$, $N_\mathrm{eng}=(3,1)$ and $N_\mathrm{eng}=(2,2)$, respectively. Different with the EMSE structure, using the ENG MMs did not significantly enlarge the total bandwidth ($\Delta \omega$) of the COWP bands because the SMs disappeared in the EMSM and the MMSM structures. On the contrary, as shown in Fig. 5(c,d), the COWP band was obviously broadened in the ENG-based MMSE structure and $\Delta \omega \approx 1.4\omega_\mathrm{p}$. That is to say, using the ENG MMs can build the ultra-broadband one-way waveguide in both the EMSE and the MMSE configurations. Moreover, as shown in Fig. 5(c), the SMs in the MMSE structure coupled with the SMPs and made a new COWP band ($\omega<\omega_\mathrm{sp}^{(1)}$). We note that the EM modes falling in the new COWP band act like regular one-way SMs, which may have potential applications such as on-chip nonreciprocal terahertz communication\cite{Gangaraj:Do,Liang:Tu}. In our opinion, using PMC wall to excite pure one-way SMs may be more practical than using ENG MMs proposed above (see Fig. 2(c)). 
	
	To verify the dispersion diagram illustrated in Fig. 5, we performed FEM simulations as working frequencies falling in the COWP bands or in the BGs, and the results are shown in Fig. 6. Symbols 'A' and 'B' represent the cases of $\varepsilon_\mathrm{m}=11.68$ and $\varepsilon_\mathrm{m}=-5$, respectively. Note that, to clearly show the difference between SMs, SMPs and the coupling modes, the magnetic field distributions instead of the electric field distributions are chosen. In Fig. 6(b), the second and the eighth pictures represent the SMs, and the sixth and the seventh pictures represent the coupling modes while the third picture represents a SMP. Interestingly, the coupling modes seem like propagate in both the medium-InSb and the metal-InSb interfaces. The other one-way propagating modes shown in Figs. 6(a) and 6(c) are SMPs, and all the results of the FEM simulations fit well with our theoretical analysis in lossless condition.
	
	Similar with the EMSE waveguide, TRT without back reflections can be achieved in the MMSE and the MMSM waveguides. As shown in Fig. 7(a), we performed the full wave simulations in a MMSE waveguide (the upper inset) which has the same parameters with the one in Fig. 4(a) except for one of the PEC wall are replaced by a layer of PMC. The lower inset indicates the electric field distribution of one of the trapped rainbow which has the operating frequency $f=0.72f_\mathrm{p}$. We further performed the FEM simulations in a MMSM waveguide which is similar with the MMSE waveguide shown in Fig.7 (a) except for the lowest layer of metal was replaced by a layer of PMC. The working frequencies in this case were considered to be $f=0.5f_\mathrm{p}$, $f=0.55f_\mathrm{p}$, $f=0.6f_\mathrm{p}$ and $f=0.65f_\mathrm{p}$, respectively. The left inset of Fig. 7(b) shows the dispersion diagram in the MMSM structures as $\varepsilon_\mathrm{m}=11.68$ (blue line),  $\varepsilon_\mathrm{m}=8$ (red line), $\varepsilon_\mathrm{m}=5$ (yellow line), $\varepsilon_\mathrm{m}=3$ (green line) and $\varepsilon_\mathrm{m}=1$ (purple line), and four horizontal arrows represent the corresponding $\omega_\mathrm{sp}^{(1)}$. It is clear that, in the MMSM structure, the COWP band limited by $\omega_\mathrm{sp}^{(1)}$ and $\omega_a$ narrows down when  $\varepsilon_\mathrm{m}$ decreases. Part of the COWP band gradually changed to a BG in the MMSM structure, in the same manner, implying the TRT. Figures 7(c)-(f) demonstrate the truly trapped rainbow with four working frequencies and the EM wave with the lower frequency was trapped at the location closer to the source. More interestingly, all the EM waves were trapped before a specific location, i.e. $z=200\upmu$m (corresponding to $\varepsilon_\mathrm{m}=0$), which implies that the TRT in a MMSM waveguide can be achieved by utilizing the epsilon-near-zero (ENZ) MMs instead of the ENG MMs. 
	
	\section{Broadband TRT and ultra-subwavelength focusing}
	\begin{figure}[ht]
		\centering\includegraphics[width=4.5 in]{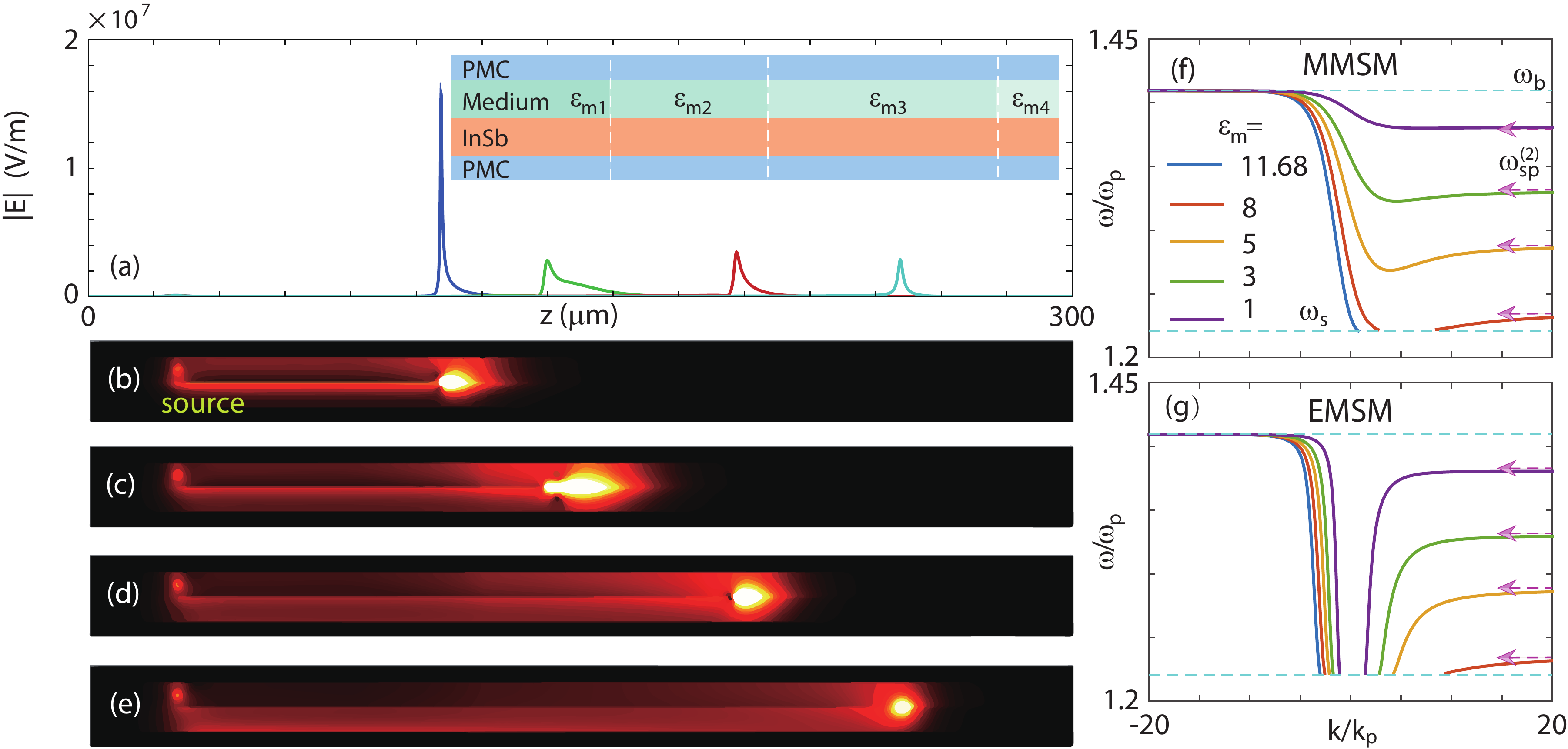}
		\caption{ Truly rainbow trapping in a MMSM waveguide consisting of ENZ MMs. (a) The electric field distributions along the interface of medium layer and InSb layer in the simulations. (b)-(e) The simulated electric field distributions as $f=1.25f_\mathrm{p}$, $f=1.3f_\mathrm{p}$, $f=1.34f_\mathrm{p}$ and $f=1.38f_\mathrm{p}$. The branches of the dispersion curves of SMPs in (f) the MMSM and (g) the EMSM waveguides when $\varepsilon_\mathrm{m}=11.68$ (red line), $\varepsilon_\mathrm{m}=8$ (blue line), $\varepsilon_\mathrm{m}=5$ (green line), $\varepsilon_\mathrm{m}=3$ (purple line) and $\varepsilon_\mathrm{m}=1$ (black line). The external magnetic field is reversed and $\omega_\mathrm{c}=-0.7\omega_\mathrm{p}$. }\label{Fig8}
	\end{figure}
	\begin{figure}[ht]
		\centering\includegraphics[width=5 in]{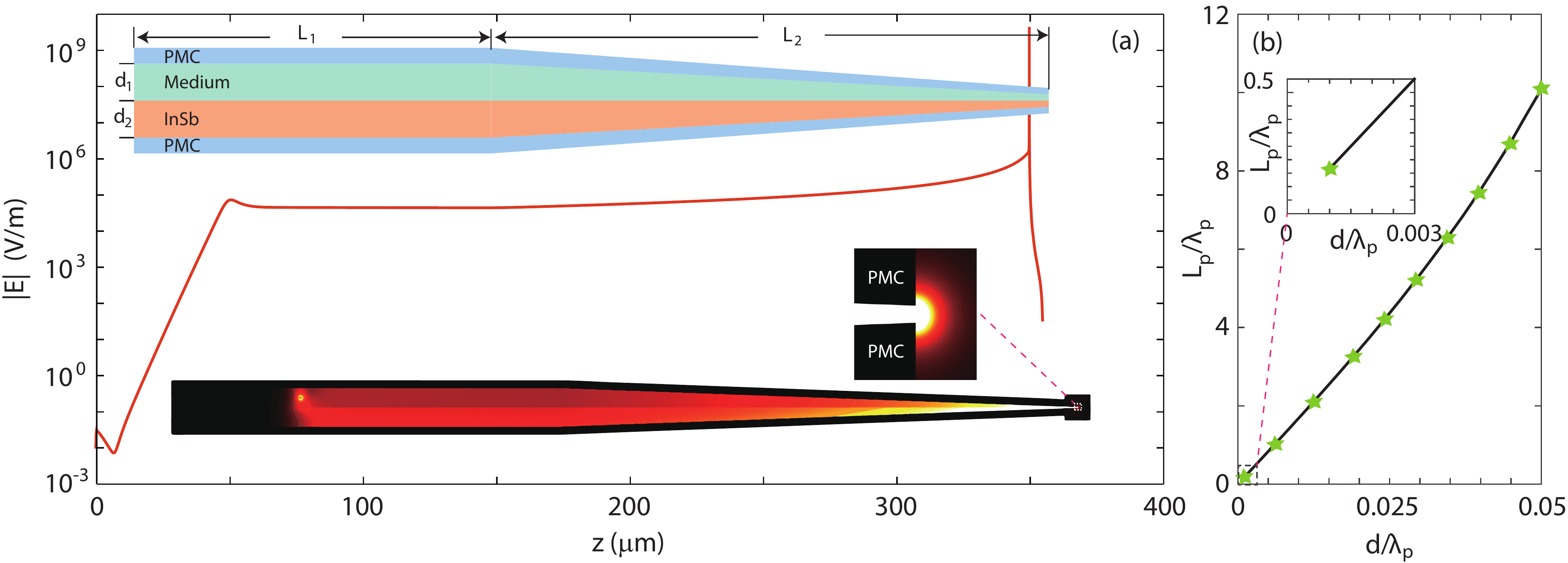}
		\caption{ (a) Ultra-subwavelength focusing based on the MMSM configuration. The upper inset shows the tapered MMSM waveguide designed for achieving ultra-subwavelength focusing. The lower inset shows the simulation result of the electric field focusing as $f=0.6f_\mathrm{p}$ and the right middle inset is the zoomed in picture around the terminal of the structure. The thicknesses of Si and InSb layers on the rightmost boundary of the structure are set to be equal and $d=0.001\lambda_\mathrm{p}$. The lengths of the uniform part and tapered part of the waveguide are respectively $L_1=150 \upmu$m, $L_2=200 \upmu$m. The other parameters are the same as in Fig. 4. (b) The propagation length ($L_\mathrm{p}$) of SMPs in the MSSM waveguide when $d_1=d_2=d$ and $0<d<0.05\lambda_\mathrm{p}$. The other parameters are $f=0.6f_\mathrm{p}$, $\omega_\mathrm{c}=0.7\omega_\mathrm{p}$ and $\nu=0.001\omega_\mathrm{p}$.}\label{Fig9}
	\end{figure}
	Besides the COWP 1 band, the COWP 2 band shown in Fig. 5(e) is also changed to a BG when $\varepsilon_\mathrm{m}$ changes from 11.68 to -5, which implies that similar TRT may be achieved in the COWP 2 band. Therefore, we performed the FEM simulations in a MMSM waveguide consisting of ENZ MMs in the COWP 2 band. The inset of Fig. 8(a) represents the ENZ-based MMSM waveguide which is similar with the one in Fig. 7(b) except for $\varepsilon_\mathrm{m1}=11.68$, $\varepsilon_\mathrm{m2}=11.68-6.68(z-L_1)/L_2$, $\varepsilon_\mathrm{m3}=5-4.99(z-(L_1+L_2))/L3$ and $\varepsilon_\mathrm{m4}=0.01$. Figures 8(b)-(e) shows the distributions of the simulated electric fields of the trapped waves. As one can see that the EM waves with four frequencies which are falling in the COWP 2 band are truly trapped in our designed MMSM waveguide since no reflections are observed in the opposite direction. We also emphasize that the EMSM model is not suitable to be used to achieve TRT in the COWP 2 band. As demonstrated in Figs. 8(f) and 8(g), with the decrease of $\varepsilon_\mathrm{m}$, the cut-off frequencies ($\omega_\mathrm{cf2}$, marked by points in Fig. 8(f)) of the dispersion curves in the MMSM structure rise up and new BG emerges while the dispersion curves in the EMSM structure rise up in $k>0$ region and unfortunately, there are SMPs in the whole region of $\omega<\omega_\mathrm{sp}^{(2)}$. Thus, we conclude that the (broadband) TRT can be achieved in two different bands in the MMSM waveguides with the total bandwidth $\Delta \omega_\mathrm{TRT} \approx 0.4\omega_\mathrm{p}$.  
	
	Due to the nonattendance of the transverse resonance constraint, our proposed structures based on PMC also can be utilized in ultra-subwavelength focusing. As shown in the inset of Fig. 9(a), a MMSM waveguide consisting of one straight part and one tapered part was proposed to study the ultra-subwavelength focusing. The medium was set to be silicon (Si) and $\varepsilon_\mathrm{m}=11.68$. The other parameters in the straight part of the MMSM structure are $L_1=150\upmu$m, $L_2=200\upmu$m and $d_1=d_2=0.05\lambda_\mathrm{p}$ ($7.5\upmu$m). The scale of right most end was set to be ultra-subwavelength and the thicknesses parameters are $d_1=d_2=0.001\lambda_\mathrm{p}$ ($0.15\upmu$m). The lowest inset of Fig. 9(a) shows the simulated electric field distribution in the designed MMSM waveguide and the working frequency $f=0.6f_\mathrm{p}$ which, according to Fig. 5(e), is in the COWP band. Note that part of the tapered part of the MMSM waveguide was set in a air box (black box in the mostright end). Moreover, in our simulations, the excited EM wave unidirectionally propagated to the end surface and no reflection or interference was observed. Figure 9(a) demonstrates the amplitude of the electric field along the Si-InSb interface and, excitingly, extremely enhanced electric field were found near the end surface. According to the simulation, the max value of the enhancement factor of the electric field is about $1.147\times 10^5$, which, as far as we know, has never been reported. To illustrate the loss impact on the electric field focusing, we calculated the propagation length ($L_\mathrm{p}$) of the SMPs with $f=0.6f_\mathrm{p}$ for $d_1=d_2=d$ and $0.001\lambda_\mathrm{p}\leq d \leq 0.05\lambda_\mathrm{p}$. As a result, $L_\mathrm{p} \approx 10.09\lambda_\mathrm{p}$ ($\approx1513.5\upmu$m) as $d=0.05\lambda_\mathrm{p}$ and $L_\mathrm{p}\approx 0.1669\lambda_\mathrm{p}$ ($\approx 25.03\upmu$m) as $d=0.001\lambda_\mathrm{p}$, which indicate that the corresponding EM wave propagates in the proposed waveguide with low loss.
	
	\section{Conclusion}
	In conclusion, we have investigated the potential uses of the epsilon negative (ENG) metamaterials (MMs) and the perfect magnetic conductor (PMC) in one-way terahertz waveguides. In our theoretical analysis, ENG MMs and PMC walls can be used to design ultra-broadband one-way waveguide since the electromagnetic modes sustained on the semiconductor-metal interface (SMs) in the ENG- or PMC-based structures become unidirectional whereas the SMs in the regular one-way waveguides always suffer from the reverse propagating SMPs. Besides, we have proposed a novel way to achieve broadband (near three times broader than the one in our previous work\cite{Xu:Sl}) truly rainbow trapping (TRT) by utilizing the gradient-index MMs. We considered three ways to use the PMC wall(s) in the metal (PEC)-medium-semiconductor-PEC (EMSE) waveguide, and three new kinds of one-way waveguides are respectively the PEC-medium-semiconductor-PMC (EMSM), the PMC-medium-semiconductor-PEC (MMSE) and the PMC-medium-semiconductor-PMC (MMSM) waveguides. Broadband TRT was theoretically investigated in the gradient-index MMSE structure consisting of ENG MMs and in the gradient-index MMSM structure consisting of epsilon-near-zero (ENZ) MMs as well. Furthermore, a straight-tapered MMSM waveguide was designed to study the ultra-subwavelength focusing at terahertz frequency. The thicknesses of the medium (silicon) and the semiconductor layers on the terminal were set to be the same and $d=0.001\lambda_\mathrm{p}$ (in this paper $\lambda_\mathrm{p}=150\upmu$m). According to the full-wave simulation, a dramatically enhanced electric field was found around the terminal and the factor of the enhancement was about $1.147\times 10^5$. The loss effect was also considered in this case and the structure is proved to be low-loss. Our proposed ultra-broadband one-way waveguides, broadband TRT theory and low-loss ultra-subwavelength focusing are promising for researches on nonlinear optics, ultra-strong electric-field devices, near-field imaging and broadband terahertz communication.
	
	\section*{Funding information}
	We acknowledge support by National Natural Science Foundation of China (NSFC) (61865009,61927813);Natural Science Foundation of Science and Tecnology Department of Sichuan Province (14JC0124,14JC0153);;Start-up funding of Southwest Medical University (20/00040186); the Science and Technology Strategic Cooperation Programs of Luzhou Municipal People's Government and Southwest Medical University (2019LZXNYDJ18).

	
	\bibliographystyle{unsrt}
	\bibliography{mybib}
	

\end{document}